\documentclass[aps,showpacs,floatfix,twocolumn,amsmath,amssymb,prd]{revtex4}
\usepackage{pstricks}

\usepackage{bm}
\usepackage{graphicx}
\usepackage{subfigure}
\usepackage{multirow}
\usepackage[citecolor=blue,colorlinks=true,urlcolor=blue,linkcolor=blue]{hyperref}
\usepackage{dsfont}  

\newcommand{\brac}[1]{\left\{#1\right\}}

\newcommand{\p}[1]{\left(#1\right)}

\newcommand{\avg}[1]{\left\langle #1\right\rangle}

\newcommand{\ba}[1]{\begin{eqnarray} #1\end{eqnarray}}

\newcommand{\be}[1]{\begin{equation} #1\end{equation}}

\newcommand{\dotp}[2]{\textbf{#1}\cdot\textbf{#2}}

\newcommand{\mr}[0]{m_{\rho}}
\newcommand{\mpi}[0]{m_{\pi}}

\newcommand{\gpp}[0]{g_{\rho\pi\pi}}
\newcommand{\ntilde}[0]{\tilde{\textbf{n}}}
\newcommand{\nhat}[0]{\hat{\textbf{n}}}
\newcommand{\tn}[0]{\tilde{\textbf{n}}}

\newcommand{\calK}[0]{\mathcal{K}}
\newcommand{\calY}[0]{\mathcal{Y}}
\newcommand{\bx}[0]{\textbf{x}}
\newcommand{\bn}[0]{\textbf{n}}

\newcommand{\bzero}[0]{\textbf{0}}
\newcommand{\calZ}[0]{\mathcal{Z}}
\newcommand{\calO}[0]{\mathcal{O}}
\newcommand{\inZ}[0]{\in\mathbb{Z}^3}

\newcommand{\bp}{\textbf{p}}
\newcommand{\Minv}{M^{-1}}

\newcommand{\beq}{\begin{equation}}
\newcommand{\eeq}{\end{equation}}

\def\av#1{\left\langle #1 \right\rangle}
\def\iden{\mathds{1}}

\newcommand{\comment}[1]{}

\begin{document}

\def\fm {\,{\rm fm}}
\def\MeV {\,{\rm MeV}}
\def\GeV {\,{\rm GeV}}

\title{Resonance parameters of the rho-meson from asymmetrical lattices}
\author{C.~Pelissier, A.~Alexandru}
\affiliation{Physics Department, The George Washington University, Washington, DC 20052, USA}

\begin{abstract}
We present a lattice QCD calculation of the parameters of the $\rho$ meson decay. The study is carried out on spatially asymmetric boxes using nHYP-smeared clover fermions with two mass-degenerate quark flavors. 
Our calculations are carried out at a pion mass $\mpi=304(2)$ MeV on the set of lattices $V=24^2\times \eta 24\times48$ with $\eta=1.0,1.25$, and 2.0 with lattice spacing $a=0.1255(7)$ fm.
The resonance mass
$\mr=827(3)(5)$ MeV and coupling constant $\gpp=6.67(42)$ are calculate using the P-wave scattering phase shifts. We construct a 2$\times$2 correlation
matrix to extract the energy of the scattering states and compute the phase shifts using the finite volume formula. By
varying the degree of asymmetry, we are able to compute a set of phase shifts that are evenly distributed throughout the spectral
region where the $\rho$ decays. 
\end{abstract}

\maketitle

\section{Introduction}

The study of multi-particle systems is an important step for lattice QCD and
our understanding of the strong force.
Experimentally, the interaction of elementary particles is 
studied using scattering methods. Due to better algorithms, more
efficient codes, and an increase in computational resources, 
it is finally possible to use lattice QCD to compute
scattering observables. Of particular interest are the scattering
channels that exhibit resonances. In this
work, we focus on the light vector meson $\rho(770)$ seen as a resonance in the elastic scattering
of two-pions in the $I^G(J^{PC})=1^+(1^{--})$ channel.

Currently, the method of choice to carry out lattice studies of scattering
phase shifts is the L\"usher method~\cite{Luscher:1985dn,Luscher:1986pf,Luscher:1990ux,Luscher:1991cf}.  
L\"usher showed that scattering phase shifts can be computed from the
spectrum of two-particle states on a torus. This method circumvents 
the ``Maiani-Testa no-go theorem" for Euclidean field theories~\cite{Maiani:1990ca}.
To measure scattering phase shifts then requires the computation of the energies
of the two-particle states. While it is known how to compute the energies
of the scattering states, the task can be quite challenging and is computationally costly. 
In the case of the $\rho$ decay, as the physical pion mass is approached 
the scattering states get closer together and we have to carefully disentangle 
the relevant two-pion states. Additionally, when the four-pion states become
dynamically relevant, the formula is no longer valid.

L\"usher's formula was developed for lattices with cubic symmetry. Two notable
extensions were worked out: (1)~systems with non-zero total momentum
where the cubic symmetry is lost due to the
relativistic boost to the center-of-mass frame~\cite{Rummukainen:1995vs} and 
(2)~systems with zero total momentum on asymmetrical lattices~\cite{Li:2003jn,Feng:2004ua}. 
The first method
can be used to investigate the $\rho$ decay on lattices where the 
zero-momentum $\rho$ cannot decay due to kinematical constraints---this was
used recently to study the $\rho$ resonance~\cite{Feng:2010es,Prelovsek:2011im,Aoki:2011yj}.
Using this method different phase shifts can be extracted by varying the total
momentum of the system. While it has been successful for studying the $\rho$ resonance,
the ability to tune the momentum of the two-pion system
is limited by the finite step size of the momenta.

In this work we chose to use the second
extension. This method 
offers a finer control over the two-pion momentum.
With a finer control over the momentum, it is possible to map out
narrower resonances, and as will be discussed later, the overall increase in computational cost
is not that significant.
Another reason for using asymmetrical lattices is that the projection onto the
relevant irreducible representation can be performed cleanly. This projection
is always performed in the center-of-mass, which coincides with the lattice frame
when the interpolators have zero momentum. For system with non-zero momentum,
the boost mixes representations and we have to rely on dynamics to disentangle 
the states of interest.

In this work we employ nHYP-smeared clover fermions with two mass-degenerate quarks.
The simulation is carried out with a pion mass $\mpi=304(2)$~MeV at a lattice spacing 
$a=0.1255(7)$~fm. We construct a $2\times2$ correlation matrix and extract the ground and 
first excited state energies
on dynamically generated ensembles of 300 gauge configurations for the three lattice volumes
\be{
V=24^2\times \eta 24\times48\;, \quad \eta=1.0,1.25,2.0\;.
}
We extract six scattering phase shifts well distributed throughout the 
resonance region and compute the resonance mass and coupling constant.
The methods used here were first tested on a set of quenched configurations
for similar quark masses and lattice sizes~\cite{Pelissier:2011ib}.

The paper is organized as follows. In Section~\ref{sec:methods} we describe the methods used and
collect the relevant formulas. In Section~\ref{sec:results} we present our results for the 
two-pion spectrum, 
compute the phase shifts, extract the resonance parameters, and compare our results with other studies. 
In Section~\ref{sec:conclusions} we present our conclusions and discuss future plans.

\section{Methods}\label{sec:methods}

\subsection{Phase shift formulas}

To compute the elastic scattering phase shifts, we use the well known
L\"usher's formula \cite{Luscher:1985dn,Luscher:1986pf,Luscher:1990ux,Luscher:1991cf}. 
In particular, we make use
of its extension to asymmetrical lattices \cite{Li:2003jn,Feng:2004ua}. In the following
we list the relevant formula needed to determine the scattering phase shifts
$\delta_{l}(k)$ for two pions with back-to-back momentum
in the angular momentum $l=1$ channel.

In this work we consider lattices with one spatial direction elongated. 
Since we work on a spatial torus, the symmetry group for zero-momentum states is reduced
from $SO(3)$ to a discrete subgroup. 
In this case the
relevant symmetry group is $D_{4h}$. For concreteness we take the $z$-direction to
be elongated.
The decomposition of the irreducible
representations of $SO(3)$ into $D_{4h}$ are listed Table \ref{tab:decomp}. From the table we
see that $l=1$ is the lowest angular momentum mode which couples to
the  $A_2^-$ and $E^-$ representations; therefore, either representation
is suitable. In the following we present the formula for these two channels.

Following~\cite{Feng:2004ua}, we introduce the generalized zeta functions
\be{
 \label{eqn:zeta_def}
 \mathcal{Z}_{lm}(s,q^2;\eta_2,\eta_3)=
 \sum_{\bn\inZ} {\mathcal{Y}_{lm}(\tilde{\textbf{n}})
 \over (\tilde{\textbf{n}}^2-q^2)^s}\;
}
where $\mathcal{Y}_{lm}(\tilde{\textbf{n}})$ are the harmonic polynomials
\be{
\mathcal{Y}_{lm}(\tilde{\textbf{n}})=\tilde n^lY_{lm}(\Omega_{\ntilde})
}
and
\be{
\ntilde = (n_1,n_2/\eta_2,n_3/\eta_3),\quad \bn\in\mathbb{Z}\;.
}
The series in Eq.~(\ref{eqn:zeta_def}) is convergent for $\mathrm{Re}\, 2s>l+3$ and can be analytically
continued to the half plane $\mathrm{Re}\, 2s > 1/2$. To compute the scattering phase shifts,
we need to evaluate $\mathcal{Z}_{lm}(s=1,q^2;\eta_2,\eta_3)$. The details are given in Appendix
\ref{sec:zeta-functions}. The parameter $q$ is related to the invariant energy $W$ of
the two-pion system through the relation
\be{
q=\frac{kL}{2\pi}\quad
}
where $k$, the pion ``momentum'', is determined by
\be{
 W=2\sqrt{\mpi^2+k^2} \,.
}
The 3D torus geometry and spatial volume is $V=L\times \eta_2 L\times \eta_3 L$. For
notational simplicity, we also introduce the quantity
\be{
\mathcal{W}_{lm}(1,q^2;\eta_2,\eta_3)=\frac{\mathcal{Z}_{lm}(1,q^2;\eta_2,\eta_3)}{\pi^{3/2}\eta_2\eta_3q^{l+1}}\,.
}
Since our lattices are elongated only in one direction, we set $\eta_2=1$ and 
$\eta_3\equiv\eta \ge 1$. The phase shift formula for the two representations
are then given by 
\ba{
\mathbf{A_2^-:}&&\cot\delta_1(k)=\mathcal{W}_{00}+\frac{2}{\sqrt{5}}\mathcal{W}_{20}\label{eqn:A-}\;,\\
\mathbf{E^-:}&&\cot\delta_1(k)=\mathcal{W}_{00}-\frac{1}{\sqrt{5}}\mathcal{W}_{20}\label{eqn:E-}\;,
}
where contributions from angular momenta $l\geq3$ are assumed to be 
negligible. 
 
We note that our formulation
is slightly different than~\cite{Feng:2004ua}, which defines the lattice volume as 
$V=\eta_1 L\times \eta_2 L\times L$.
The phase shift formulas in either case are equivalent.
Additionally, our formula for the $E^-$ channel differs by the exclusion of the term 
$\pm \sqrt{3/10}(\mathcal{W}_{22}+\mathcal{W}_{2-2})$, which vanishes due to symmetry under rotations
around the elongated direction.

Lastly, we consider the case of a cubic lattice. The symmetry group is $O_h$, and the relevant representation
is $F_1^-$ (see Table \ref{tab:decomp}). In this case $\mathcal{W}_{20}$ vanishes, 
Eq.~(\ref{eqn:A-}) and Eq.~(\ref{eqn:E-}) become equivalent,
and the phase shift formula is given by setting the right hand side equal to $\mathcal{W}_{00}$.
\begin{table}
\caption{Resolution of the $2J+1$ spherical harmonics into the irreducible representations of $O_h$ and $D_{4h}$.}
\label{tab:decomp}
\center
\begin{ruledtabular}
\begin{tabular}{ccc}
J&$O_h$ & $D_{4h}$ \\
\hline
0&$A^+_1$&$A^+_1$\\
1& $F^-_1$& $A^-_2\oplus E^-$\\
2& $E^+\oplus F^+_2$&$A^+_1\oplus B^+_1\oplus B^+_2 \oplus E^+$\\
3& $A^-_2 \oplus F^-_1\oplus F^-_2$& $A^-_2\oplus B^-_1\oplus B^-_2 \oplus 2E^-$\\
4& $A^+_1 \oplus E^+ \oplus F^+_1\oplus F^+_2$ &$2A^+_1 \oplus A^+_2\oplus B^+_1 \oplus B^+_2\oplus 2E^+$\\
\end{tabular}
\end{ruledtabular}
\end{table}

\subsection{Selecting lattices}

In order to extract the resonance parameters for $\rho$, we need to compute the scattering phase shifts
in the spectral region where the resonance appears. 
To minimize statistical errors, it is preferable to compute the phase
shifts using the lowest lying states in the channel. To achieve this, we have to select appropriate
lattice volumes and, for practical purposes, make a selection which is computationally economic. This choice
is difficult since the resonance mass and coupling constant are generally not known \textit{a priori} for
unphysical pion masses. However, using recent lattice results~\cite{Feng:2010es,Lang:2011mn,Aoki:2011yj}
and predictions from unitarized chiral perturbation theory~\cite{Pelaez:2010fj}, 
a reasonably good guess can be made. In the following we describe 
the method we used to select lattices.

\begin{figure}[b!]
\centering
\includegraphics[width=3.25in]{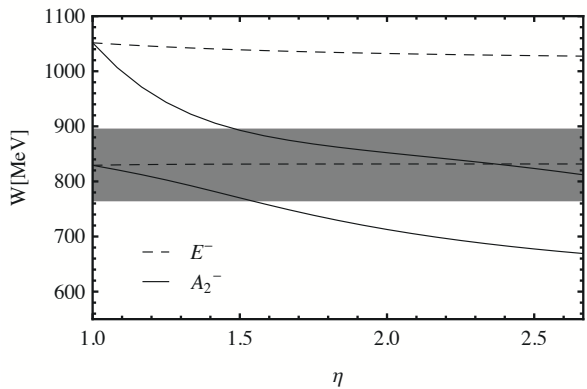}
\caption{Estimated spectrum for $\mpi=300$ MeV, $\gpp=6$, and $\mr=830$ MeV.
The lattice size is set to $L=3$ fm.
The two curves indicate the ground, and first excited states.
The shaded region is the resonance region, i.e., $\mr\pm\Gamma_\rho$. }
\label{fig:est-spectrum}
\end{figure}

To estimate the two-pion energies, we use the effective range formula \cite{Nakamura:2010zzi}
\begin{equation}
\cot{\delta}_1(W) = \frac{6\pi}{g^2_{\rho\pi\pi}} \frac{W(m_{\rho}^2 - W^2)}{\p{\frac{W^2}{4}-m_{\pi}^2}^{3/2}}\,,
\label{eqn:eff-formula}
\end{equation}
which is known to parameterized the $\rho$ resonance well at the physical pion mass. Using the equation above together with Eqs.~(\ref{eqn:A-})~and~(\ref{eqn:E-}),
we can compute the expected spectrum as a function of $\eta$ provided we can estimate the values of $\mr$ and $\gpp$.
From recent lattice studies we know that the coupling constant has little pion mass dependence and 
is close to the physical value $\gpp\approx 6$ for pion masses 
$\mpi < 350$ MeV~\cite{Feng:2010es,Lang:2011mn,Aoki:2011yj}. As a result,
we start with the assumption that $\gpp=6$. To estimate the value of $\mr$, we use the result from
unitarized chiral perturbation theory which shows reasonable agreement with the lattice
results~\cite{Pelaez:2010fj}. Note that this picture can be refined as we collect data
from lattice simulations. When choosing subsequent lattices, we use estimates
for $\mr$ and $\gpp$ based on the available lattice data.

In Fig.~\ref{fig:est-spectrum} we plot the expected spectrum
for a pion mass of $300$ MeV. Note that the two-dimensional $E^{-}$ representation
corresponds to the two states with non-zero back-to-back momentum in the transversal directions,
whereas the $A_2^{-}$ is the one-dimensional representation corresponding to
the state with the momentum along the longitudinal/elongated direction.
For the $E^-$ representation the spectrum depends very little on $\eta$. 
The reason is that the relative momentum of the two-pion state depends on the size of
the transversal direction, which does not change. Thus, computing the energy of the states in
the $E^-$ representation will produce phase shifts very similar to the ones on a cubic lattice.
The $A_2^-$ representation shows a strong dependence on the
value of $\eta$ and offers good control over the value of the ground and first excited state energies. To
generate additional phase shifts, one then needs to work in the $A_2^-$ channel and select appropriate values of
$\eta$. We see that by choosing $\eta=1$, $1.25$, and $2.0$, we should get
a good scan of the resonance region. 

\comment{
Since we are interested in the $l=1$ channel, we make use of the interpolators
$\calO_i$ with $i=1,2,$ and 3 which transform under rotations as
\be{
U(R)\calO_iU^{\dag}(R)=\sum_jD(R)_{ij}\calO_j
}
where $D(R)$ are the representation matrices for the coordinate basis $x_i$. If we consider
the case where $\eta_2\neq\eta_3\neq1$, we can project onto 
the set of two-pion states with back-to-back momentum in the $i^{th}$ spatial-direction 
by using the interpolator $\calO_i$ which couples to one the three one-dimensional
representations of $D_{2h}$. In this case, the momentum of the two pions
will be
\be{
\bp_i = \frac{2\pi}{\eta_i L}n_i\,\hat{i}\;,\quad n_i \in \mathbb{Z}/\{0\}\;,
}
and the two-pion spectrum will be non-degenerate. One can also consider the cases where
$\eta_3\neq\eta_2=1$ and $\eta_3=\eta_2=1$ in which case the relevant rotation groups
are $D_{4h}$ and $O_h$, respectively. In Table \ref{tab:2pimom}, we list the the momentum
of the two least energetic two-pion states for the three cases.
From these remarks, it is clear that we have good control over the momentum of the two pions.
However, the energy of the two-pion system also depends on the interaction and the size and shape of the lattice.
In order to estimate the value of the two-pion energies, in particular the ground and first excited state, requires
additional input.
}

Lastly, we make a few remarks about our choice to use lattices with a single direction elongated. 
It may seem that a better choice for these types of studies are orthorhombic lattices, i.e., 
lattices in which all three side have different lengths. In this case 
six different phase shifts can be computed from one ensemble using only the ground 
and first excited state energies. Choosing appropriate values for 
$L,\eta_2,$ and $\eta_3$, we could get all of these phase shifts in the resonance region. 
Since the computational effort scales with $V^{5/4}$~\cite{Ukawa:2002pc},
in this case the computational cost is $(\eta_2\eta_3)^{5/4}\times\tau$ 
where $\tau$ denotes the cost for the cubic lattice. If instead we choose a lattice with 
only one direction elongated
and generate the two ensembles with $\eta=\eta_2$ and $\eta=\eta_3$, the cost is 
$(\eta_2^{5/4}+\eta_3^{5/4})\times \tau$. The overall computational cost of the two choices is 
similar when the elongation parameters $\eta_{2,3}$ are small. Additionally, using two 
different ensembles reduces the correlation
between phase shifts and allows for studies of finite volume effects of other observables. 
In light of these considerations, we choose to generate lattices with only a single direction
elongated.

\subsection{Variational analysis}

To compute the low lying two-pion energies with the quantum numbers $I^G(J^{PC})=1^+(1^{--})$ of the
$\rho$, we employ the variational method proposed by L\"uscher
and Wolff \cite{Luscher:1990ck}. We use a two-dimensional variational basis and construct the
correlation matrix
\be{
C(t)_{ij}=\langle\calO_i(t)\calO_j^{\dag}(0)\rangle\;
}
to extract
the ground and first excited states. We compute the eigenvalues of
\be{
C(t_0)^{-1/2}C(t)C(t_0)^{-1/2}\psi^{(n)}(t,t_0)=\lambda^{(n)}(t,t_0)\psi^{(n)}(t,t_0)
}
for each time slice $t$. The two-pion energies are then determined from the
long-time behavior of the eigenvalues~\cite{Blossier:2009kd}
\be{
\lambda^{(n)}(t,t_0)\propto e^{-E_nt}(1+\mathcal{O}(e^{-\Delta E_n t}))\,,\quad n=1,2
}
where $\Delta E_n=E_3-E_n$. For the variational basis we use the interpolators
\ba{
\calO_1&=&\frac{1}{\sqrt{2}}\brac{\pi^+(\bp_z)\pi^-(-\bp_z)-\pi^-(\bp_z)\pi^+(-\bp_z)}\;,\\
\calO_2&=&\sum_{\bx}\frac{1}{\sqrt{2}}\brac{\bar{u}(x)\gamma_3u(x)-\bar{d}(x)\gamma_3d(x)}
} 
where $\bp_z = \frac{2\pi}{\eta L}\hat{\bm z}$ is the lowest lattice momentum in the elongated
direction and 
\ba{
\pi^{-}(\bp)&=&\sum_\bx\bar{u}(\bx)\gamma_5d(\bx)e^{i\bp\bx}\,,\\
\pi^{+}(\bp)&=&\sum_\bx\bar{d}(\bx)\gamma_5u(\bx)e^{i\bp\bx}\,.
}
Lastly, we note that as defined above $\calO_1$ and $\calO_2$
are hermitian and anti-hermitian, respectively.

\subsection{Evaluation of the correlation matrix}

%
\def\square#1 #2 #3{%
\psset{linewidth=1pt}
\if1#1{\psline{->}(0,1)(0.5,1)\psline(0.5,1)(1,1)}\fi
\if2#1{\psline(0,1)(0.5,1)\psline{<-}(0.5,1)(1,1)}\fi
\if3#1{\psline(0,1)(1,1)}\fi
\if1#2
\psline(1,1)(1,0)
\psline(1,0)(0,0)
\psline(0,0)(0,1)
\else
\psline(1,1)(0,0)
\psline(0,0)(1,0)
\psline(1,0)(0,1)
\fi
\if1#3
\rput[t](-.3,.1){$\substack{\gamma_5\\(-\bm p, t_i)}$} 
\rput[t](1.3,.1){$\substack{\gamma_5\\(\bm p, t_i)}$} 
\rput[b](-.3,.9){$\substack{(-\bm p, t_f)\\\gamma_5}$} 
\rput[b](1.3,.9){$\substack{(\bm p, t_f)\\\gamma_5}$} 
\fi
}
\def\pipicross{%
\psset{linewidth=1pt}
\pscurve(0,1)(0.35,0.4)(1,0)
\pscurve(1,0)(0.65,0.6)(0,1)
\pscurve(1,1)(0.65,0.4)(0,0)
\pscurve(0,0)(0.35,0.6)(1,1)
\psline{->}(0.69,0.45)(0.7,0.46)
\psline{->}(0.6,0.65)(0.59,0.66)
}
\def\pionprop{%
\psset{linewidth=1pt}
\pscurve(0.25,1)(0.4,0.5)(0.25,0)
\pscurve(0.25,1)(0.1,0.5)(0.25,0)
\psline{->}(0.1,0.42)(0.1,0.41)
}
\def\pipidir{%
\pionprop
\rput(0.5,0)\pionprop
}
\def\triangle#1 #2 #3{%
\psset{linewidth=1pt}
\if1#2
\def\up{1}\def\down{0}
\else
\def\up{0}\def\down{1}
\fi
\rput(0,\up){%
\if1#1{\psline{->}(0,0)(0.5,0)\psline(0.5,0)(1,0)}\fi
\if2#1{\psline(0,0)(0.5,0)\psline{<-}(0.5,0)(1,0)}\fi
\if3#1{\psline(0,0)(1,0)}\fi}
\psline(0.5,\down)(0,\up)
\psline(0.5,\down)(1,\up)
\if1#3
\if2#2
\rput[t](-.3,.1){$\substack{\gamma_5\\(-\bm p, t_i)}$} 
\rput[t](1.3,.1){$\substack{\gamma_5\\(\bm p, t_i)}$} 
\rput[b](.5,1.1){$\substack{(\bm 0, t_f)\\\gamma_3}$}
\else
\rput[b](-.3,.9){$\substack{(-\bm p, t_f)\\\gamma_5}$} 
\rput[b](1.3,.9){$\substack{(\bm p, t_f)\\\gamma_5}$} 
\rput[t](.5,-.1){$\substack{\gamma_3\\(\bm 0, t_i)}$}
\fi
\fi
}

To evaluate the correlation matrix, one needs to compute the following diagrams:
\be{
\begin{pspicture}(6.3,2.8)
\psset{unit=0.7cm}
\rput(0.6,3.9){$C_{11}(t)$}
\psline[linewidth=0.5pt](0.,3.5)(8.5,3.5)
\rput(0,2.25){\square1 1 0}
\rput(1.5,2.25){\square2 1 0}
\rput(3,2.25){\square1 2 0}
\rput(4.5,2.25){\square2 2 0}
\rput(6,2.25){\pipicross}
\rput(7.5,2.25)\pipidir

\rput(0.6,1.65){$C_{12}(t)$}
\psline[linewidth=0.5pt](0.,1.25)(8.5,1.25)
\rput(0,0.0){\triangle1 1 0}
\rput(1.5,0.0){\triangle2 1 0}

\rput(3.75,1.65){$C_{21}(t)$}
\rput(3.25,0.0){\triangle1 2 0}
\rput(4.75,0.0){\triangle2 2 0}

\rput(7,1.65){$C_{22}(t)$}
\rput(6.75,0.0)\pionprop
\end{pspicture}
\nonumber}
Above, we used a compressed notation; more explicitly, we have
\ba{
\begin{pspicture}(-0.25, 0.25)(.8,1.1)
\psset{unit=.7cm}
{\square1 1 1}
\end{pspicture} &=& {\rm Tr\,}[ e^{-i\bm p}\gamma_5 M^{-1}(t_i, t_i) e^{i\bm p} 
\gamma_5 M^{-1}(t_i, t_f) \times\nonumber\\
&&\,\, e^{i\bm p} \gamma_5 M^{-1}(t_f, t_f) e^{-i\bm p} \gamma_5 M^{-1}(t_f, t_i)] \,,
}
where the square matrices
\ba{
(\gamma_5)_{\bm x, a, \alpha; \bm y, b, \beta} &\equiv& 
\delta(\bm x-\bm y) (\gamma_5)_{ab} \delta_{\alpha\beta} \,,\nonumber\\
(e^{i\bm p})_{\bm x, a, \alpha;\bm y, b,\beta} &\equiv& 
e^{i\bm p\bm x}\delta(\bm x-\bm y)\delta_{ab} \delta_{\alpha\beta} \,,\\
M^{-1}(t_f,t_i)_{\bm x, a, \alpha; \bm y, b, \beta} &\equiv&
M^{-1}(\bm x,t_f, a, \alpha; \bm y, t_i, b, \beta) \,,\nonumber
} 
have $12\times N_s$ rows with $N_s$ the number of points in a time slice.
Above, $M$ is the quark matrix and thus $M^{-1}$ is the regular quark propagator.
All four point diagrams are defined in a similar manner. The diagrams with
three points can be written using the following template:
\ba{
\psset{unit=.7cm}
\pspicture[shift=-0.5](-.2, -.0)(1.5,1.9)
{\triangle2 2 1}
\endpspicture
&=& {\rm Tr\,}[ e^{-i\bm p}\gamma_5 M^{-1}(t_i, t_i)\times\nonumber\\
&&\,\, e^{i\bm p}\gamma_5 M^{-1}(t_i,t_f) \gamma_3 M^{-1}(t_f, t_i)] \,.
}

Using the $\gamma_5$-hermiticity of the quark propagator, i.e., 
$M^{-1}(t,t')^\dagger=\gamma_5 M^{-1}(t',t)\gamma_5$, we can show that
the four-point diagrams are purely real and the three-point diagrams are purely
imaginary, configuration by configuration. Moreover, using parity 
we can show that
\begin{align}
\left\langle
\psset{unit=0.6cm}
\begin{pspicture}[shift=-0.35](-.1,-.1)(1.1,1.1)
{\triangle1 2 0}
\end{pspicture}
\right\rangle_U &=-
\left\langle
\psset{unit=0.6cm} 
\begin{pspicture}[shift=-0.35](-.1,-.1)(1.1,1.1)
{\triangle2 2 0}
\end{pspicture}
\right\rangle_U
,&\,
\left\langle
\psset{unit=0.6cm}
\begin{pspicture}[shift=-0.5](-.1,-.1)(1.1,1.1)
{\triangle1 1 0}
\end{pspicture}
\right\rangle_U &=-
\left\langle
\psset{unit=0.6cm} 
\begin{pspicture}[shift=-0.5](-.1,-.1)(1.1,1.1)
{\triangle2 1 0}
\end{pspicture}
\right\rangle_U
,\nonumber\\
\left\langle
\psset{unit=0.6cm}
\begin{pspicture}[shift=-0.5](-.1,-.1)(1.1,1.1)
{\square1 1 0}
\end{pspicture}
\right\rangle_U &=
\left\langle
\psset{unit=0.6cm} 
\begin{pspicture}[shift=-0.5](-.1,-.1)(1.1,1.1)
{\square2 1 0}
\end{pspicture}
\right\rangle_U
,&\,
\left\langle
\psset{unit=0.6cm}
\begin{pspicture}[shift=-0.5](-.1,-.1)(1.1,1.1)
{\square1 2 0}
\end{pspicture}
\right\rangle_U &=
\left\langle
\psset{unit=0.6cm} 
\begin{pspicture}[shift=-0.5](-.1,-.1)(1.1,1.1)
{\square2 2 0}
\end{pspicture}
\right\rangle_U.
\end{align}
Time reversal symmetry allows us to connect the quark diagrams 
required for $C_{12}$ with the ones for $C_{21}$. In particular,
one can show that
\be{%
\left\langle
\psset{unit=0.6cm} 
\begin{pspicture}[shift=-0.5](-.1,-.1)(1.1,1.1)
{\triangle2 1 0}
\end{pspicture}
\right\rangle_U=
\left\langle
\psset{unit=0.6cm}
\begin{pspicture}[shift=-0.35](-.1,-.1)(1.1,1.1)
{\triangle2 2 0}
\end{pspicture}
\right\rangle_U\,.
}
Above, we use $\left\langle\cdot\right\rangle_U$ to denote the average
with respect to the gauge configurations. All together, the four
components of the correlation matrix can be constructed from
\ba{
C_{11}(t) &=& \left\langle
2\psset{unit=0.6cm}
\begin{pspicture}[shift=-0.4](-.2,-.1)(1.1,1.1)
{\square1 1 0}
\end{pspicture}
-2\psset{unit=0.6cm}
\begin{pspicture}[shift=-0.4](-.2,-.1)(1.1,1.1)
{\square1 2 0}
\end{pspicture}
+
\psset{unit=0.6cm}
\begin{pspicture}[shift=-0.4](-.2,-.1)(1.1,1.1)
{\pipicross}
\end{pspicture}
-
\psset{unit=0.6cm}
\begin{pspicture}[shift=-0.4](-.2,-.1)(1.1,1.1)
{\pipidir}
\end{pspicture}
\right\rangle_U\,,\nonumber\\
C_{12}(t) &=& - C_{21}(t) = 
\left\langle
\psset{unit=0.6cm} 
\begin{pspicture}[shift=-0.5](-.1,-.1)(1.1,1.1)
{\triangle2 1 0}
\end{pspicture}
\right\rangle_U \,,\\
C_{22}(t) &=&
\left\langle
\psset{unit=0.6cm} 
\begin{pspicture}[shift=-0.45](-.1,-.1)(.6,1.1)
{\pionprop}
\end{pspicture}
\right\rangle_U\,.\nonumber
}

Note that $C_{12}$ and $C_{21}$ are related. To save time,
we compute only $C_{21}$ and use it to determine $C_{12}$.
$C_{22}$ can be evaluated using standard methods but
the terms for $C_{11}$ and $C_{21}$ require the 
all-to-all propagator. As a result, they must be computed stochastically.
Our calculation follows the steps described in a study by the
CP-PACS collaboration~\cite{Aoki:2011yj}. 
We introduce a random $Z(4)$ noise time-slice vector $\xi$ (with color and
spinor components), satisfying the condition
\be{
\av{\xi\xi^\dagger}_\xi = \iden\,
}
where $\iden$ is the identity matrix in $12\times N_s$ dimensions
and $\av{\cdot}_\xi$ denotes the average over the noise.
Using these noise vectors, we define the time-slice vectors
\ba{
u(t | \bm p, t_2,\xi) &=& \Minv(t,t_2) e^{i\bm p}\xi
\label{eqn:u}\\
v(t | \bm p_1, t_1 | \bm p_2, t_2, \xi) &=&
\Minv(t, t_1) e^{i\bm p_1}\gamma_5
u(t_1 | \bm p , t_2, \xi)\,.
\nonumber \label{eqn:v}
} 
With the above definitions unbiased estimators for the quark diagrams for $C_{11}(t)$ and $C_{21}(t)$ 
can be constructed as
\ba{
\psset{unit=0.6cm}
\begin{pspicture}[shift=-0.4](-.2,-.1)(1.1,1.1)
{\square1 1 0}
\end{pspicture} &=& \av{v(t_f|\bm p,t_i|-\bm p,t_i,\xi)^\dagger e^{-i\bm p}
v(t_f|\bm p,t_f|\bm 0,t_i,\xi)}_{\xi,U} \,,
\nonumber\\
\psset{unit=0.6cm}
\begin{pspicture}[shift=-0.4](-.2,-.1)(1.1,1.1)
{\square1 2 0}
\end{pspicture} &=& \av{v(t_f|-\bm p,t_i|\bm p,t_i,\xi)^\dagger e^{-i\bm p}
v(t_f|\bm p,t_i|\bm 0,t_i,\xi)}_{\xi,U} \,,
\nonumber\\
\psset{unit=0.6cm}
\begin{pspicture}[shift=-0.4](-.2,-.1)(1.1,1.1)
{\pipicross}
\end{pspicture} &=& \Big\langle\av{u(t_f|\bm 0,t_i, \xi)^\dagger e^{-i\bm p}
u(t_f|\bm p,t_i,\xi)}_\xi \times 
\nonumber\\
&&\quad\av{u(t_f|\bm 0,t_i, \eta)^\dagger e^{i\bm p}
u(t_f|-\bm p,t_i,\eta)}_\eta \Big\rangle_U \,,
\nonumber\\
\psset{unit=0.6cm}
\begin{pspicture}[shift=-0.4](-.2,-.1)(1.1,1.1)
{\pipidir}
\end{pspicture} &=& \Big\langle\av{u(t_f|\bm 0,t_i, \xi)^\dagger e^{-i\bm p}
u(t_f|-\bm p,t_i,\xi)}_\xi \times 
\nonumber\\
&&\quad\av{u(t_f|\bm 0,t_i, \eta)^\dagger e^{i\bm p}
u(t_f|\bm p,t_i,\eta)}_\eta \Big\rangle_U \,,
\\
\psset{unit=0.6cm}
\begin{pspicture}[shift=-0.35](-.2,-.1)(1.1,1.1)
{\triangle2 2 0}
\end{pspicture} &=& \av{u(t_f|\bm 0, t_i, \xi)^\dagger \gamma_5\gamma_3 
v(t_f|-\bm p,t_i|\bm p,t_i,\xi)}_{\xi,U} \,,
\nonumber\\
\psset{unit=0.6cm}
\begin{pspicture}[shift=-0.35](-.2,-.1)(1.1,1.1)
{\triangle1 2 0}
\end{pspicture} &=& \av{u(t_f|\bm 0, t_i, \xi)^\dagger \gamma_5\gamma_3
v(t_f|\bm p,t_i|-\bm p,t_i,\xi)}_{\xi,U} \,.
\nonumber
}

\begin{figure*}[!th]
\includegraphics[width=3.25in]{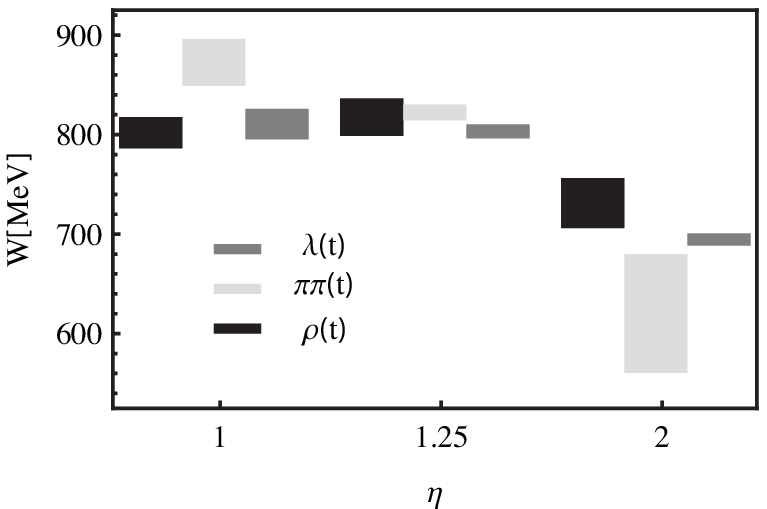}
\includegraphics[width=3.25in]{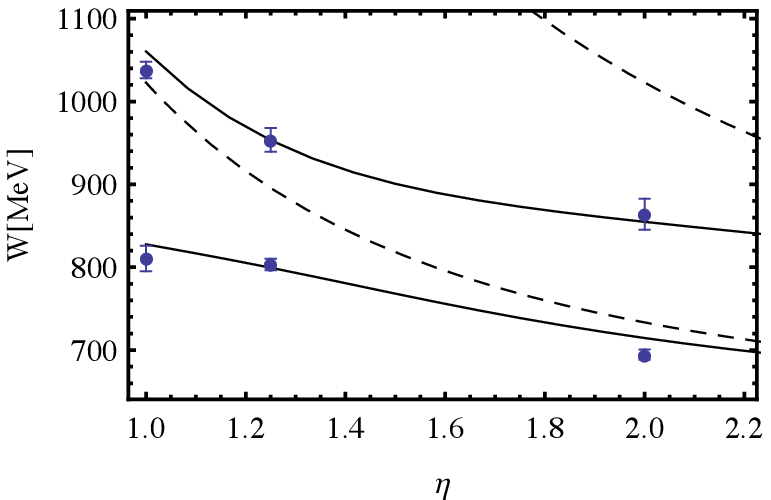}
\caption{Left panel: ground state energies for the three different elongations 
extracted using different methods. Right panel:
ground and first excited state energies compared to the effective range formula
using the parameters extracted from our fits. The dashed lines represent the
expected two-pion energies in the absence of interaction, i.e., 
$W_n=2\sqrt{m_\pi^2+(2\pi n/\eta L)^2}$.}
\label{fig:specplots}
\end{figure*}

Lastly, we make a few comments on the variance of the stochastic estimators. As is
common practice, we chose to use spin-color dilution. To check its effectiveness,
we compared the variance of $C_{11}(t)$ and $C_{21}(t)$ with and without dilution 
in the temporal range where the fitting is performed. Using spin-color dilution, 
we found an order of magnitude decrease in the
variance for the same numerical effort. Another decision which has to
be made is the size of the noise ensembles. 
Using a simple model, we can show that the overall variance of the estimator is
\be{
\sigma^2\approx\frac{\sigma_U^2}{N}\p{1+\frac{\sigma^2_\xi/\sigma_U^2}{M}}
}
where $\sigma_U^2$ and $\sigma_\xi^2$ are the contributions to the overall variance
from the gauge and noise fluctuations, $N$ is the number of gauge configurations, and $M$ 
the number of noises per configuration.
In the temporal region where the fits
were performed, we found that the ratio $\sigma^2_\xi/\sigma_U^2$ 
is small and that the variance is dominated by gauge fluctuations.
As a result, the variance is not significantly decreased even
after a single noise. This is similar to the results found by \cite{Aoki:2011yj} where
no substantial decrease in the variance was found after two noises. In this work,
we also chose to use two noises. 

The stochastic estimator requires thousands of inversions for 
each configuration making the numerical cost of the analysis 
comparable with the cost of generating the dynamical configurations. 
To carry out these calculations, we used our own GPU BiCGstab 
inverters~\cite{Alexandru:2011ee}.

\section{Results}\label{sec:results}

\subsection{Two-pion spectrum}

To compute the two-pion spectrum, we generated three dynamical ensembles of
300 gauge configurations. We employed the L\"usher-Weiss
action with $\beta=7.1,\, \kappa=0.1282$, and $u_0=0.868$. The simulation was
carried out using nHYP-smeared clover fermions with two mass-degenerated quarks and
the standard smearing parameters 
$\alpha_1=0.75\,,\alpha_2=0.6\,,$ and $\alpha_3=0.3$~\cite{Hasenfratz:2007rf}.
The spatial volume of the the three lattices was
\be{
24\times24\times \eta 24\,, \quad \eta=1.0\,,1.25,2.0\,,
}
with temporal extent $N_t=48$. The lattice spacing for all three ensembles 
was determined to be $a=0.1255(7)$ using the Sommer scale 
(see Appendix~\ref{app:lattice_spacing}), and
the computed value for the pion mass was $\mpi=304(2)$ MeV.

\begin{figure*}[!t]
\includegraphics[width=3.25in]{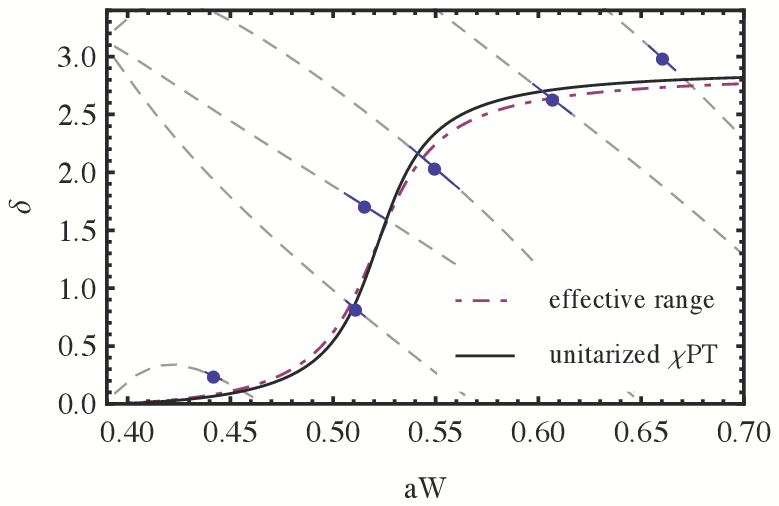}
\includegraphics[width=3.25in]{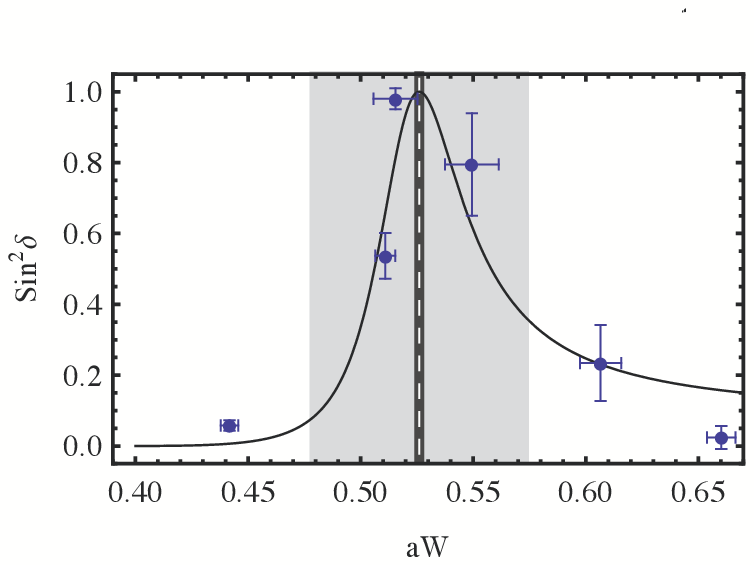}
\caption{Left panel: scattering phase shifts computed in this study, effective
range formula fit, and the results from unitarized $\chi$PT~\cite{Pelaez:2010fj}. 
The dashed lines describe the energy dependence of the phase shift according to
Eq.~(\ref{eqn:A-}). Right panel: cross section as a function of the center-of-mass
energy. The dashed line indicates the $\rho$ mass, the narrow dark band its error, and the 
shadowed box indicates the resonance region $\mr\pm\Gamma$.}
\label{fig:phaseplots}
\end{figure*}

We only computed the energies in the $A_2^-$ channel since the energies in the $E^-$
channel are not significantly different than those obtained in the $F^-$ 
channel on the cubic lattice. The numerical values are listed in 
Table~\ref{tab:dyn-spec}. In the left panel of Fig.~\ref{fig:specplots}, we show 
the ground state energies obtained using solely the $\pi\pi$ and $\rho$ interpolators, 
as well as the energy obtained using the variational method. The results 
show consistent behavior as the lattice size is increased with the energy and, as expected,
decreases as the back-to-back momentum of the pions becomes smaller. Note that the individual
correlators, extracted separately from $C_{11}(t)$ and $C_{22}(t)$, have larger
error bars. For our fits we use the energies extracted using the variational
method.
In the right panel of Fig.~\ref{fig:specplots}, we show our results for the
ground and first excited states, as well as the spectral behavior as 
predicted by the effective
range formula Eq.~(\ref{eqn:eff-formula}) using our fit results.  
We see that the effective range formula describes the lattice data well
and that the lattice data resolves clearly the effect of the
pions' interaction.

\begin{table}[!b]
\caption{Two pion energies in then $A_2^-$ sector.}
\label{tab:dyn-spec}
\center
\begin{ruledtabular}
\begin{tabular}{cccc}
$\eta$&1.0&1.25&2.0\\
\hline
$aE_0$&0.516(10)&0.511(4)&0.442(4)\\
$aE_1$&0.660(6)&0.606(10)&0.549(12)\\
$\delta_1(E_0)$&1.71(11)&0.822(65)&0.242(31)\\
$\delta_1(E_1)$&2.99(10)&2.64(13)&2.04(18)\\
$a\mpi$&0.1925(7)&0.1944(6)&0.1946(8)\\
\end{tabular}
\end{ruledtabular}
\end{table}

\subsection{Phase shifts and resonance parameters}

To compute the resonance parameters, we evaluate the scattering formula Eq.~(\ref{eqn:A-}) with
the energies obtained for the ground and first excited states. The resonance mass and coupling
constant are then computed by fitting the scattering phase shifts with the effective range
formula given in Eq.~(\ref{eqn:eff-formula}). We use a correlated $\chi^2$-fit 
since the energies for the ground and first excited states are extracted from the same ensemble
and construct the covariance matrix with two-by-two blocks, one per ensemble.

Note that the standard $\chi^2$ method needs to be modified for our fit: the
statistical errors affect both $x$ and $y$ coordinates in our plot since the 
phase shifts are related to the energy. 
Referring to the left panel of Fig.~\ref{fig:phaseplots}, as the energy is varied the phase shifts
move along the dashed lines indicated in the figure. The solid lines indicate the 
error bars obtained by varying the energy in the $W\pm\sigma_W$ interval.
The $\chi^2$ function is defined to be
\be{
\chi^2=\frac{1}{2}{\Delta}^TC^{-1}{\Delta}
}
where $\Delta$ is the difference vector
\be{
\Delta_i=W_i-W_i^0\,,\label{eqn:diffvec}
}
$W_i$ indicates the computed value, and $W_i^0$ the value at the intersection of the dashed
line and the fit curve. 

The decay width is determined using the fit
results and the relation
\be{
\Gamma =\frac{g_{\rho\pi\pi}^2}{6\pi}\frac{\p{m_{\rho}^2/4-m_{\pi}^2}^{3/2}}{m_{\rho}^2}\;.\label{eqn:width}
}
We find
\be{
\begin{aligned}
&\gpp = 6.67(42)\,,\quad &\mr&= 827(3)(5)\;\rm{MeV}\,,\\
&\Gamma=76.6(20)(5)\;\rm{MeV}\,,\quad&\Gamma_{ph}&=184(23)\;\rm{MeV}\,,
\end{aligned}
}
where $\Gamma_{ph}$ is computed using the computed value of $\gpp$ and the physical values
of $\mr$ and $\mpi$. The first quoted uncertainty indicates the statistical uncertainty 
and the second indicates the uncertainty associated with the determination
of the lattice spacing.

In Fig.~\ref{fig:phaseplots} we show the scattering phase shifts and the fit curve we obtained
using the effective range formula. With the exception of the left-most and right-most points, 
the data are not more than one standard deviation away from the curve. 
However, the confidence level of the fit is $Q\approx 8\%$, and it is not
clear whether the effective range formula is reliable for $\mpi=304(2)$~MeV. Assuming the
effective range formula is reliable, the value we computed for the coupling constant is in
reasonably good agreement with the physical value quoted by the PDG $\gpp^{phys}=5.975(16)$~\cite{Nakamura:2010zzi}.

\begin{figure*}[!t]
\includegraphics[width=3.25in]{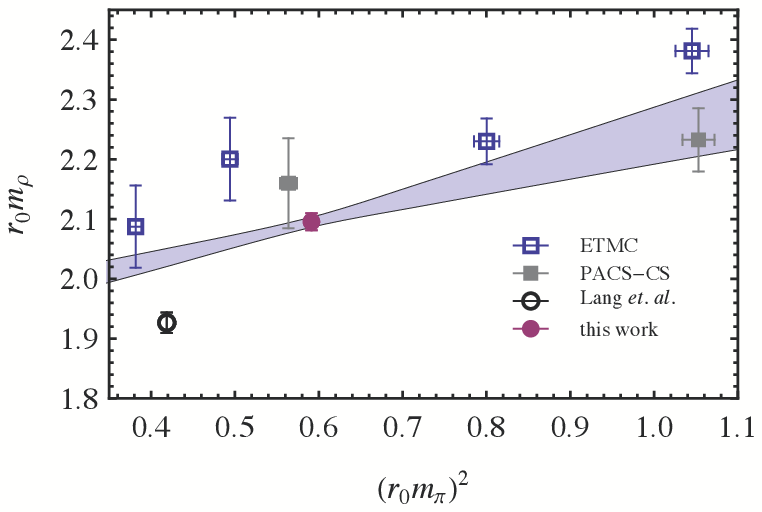}
\includegraphics[width=3.25in]{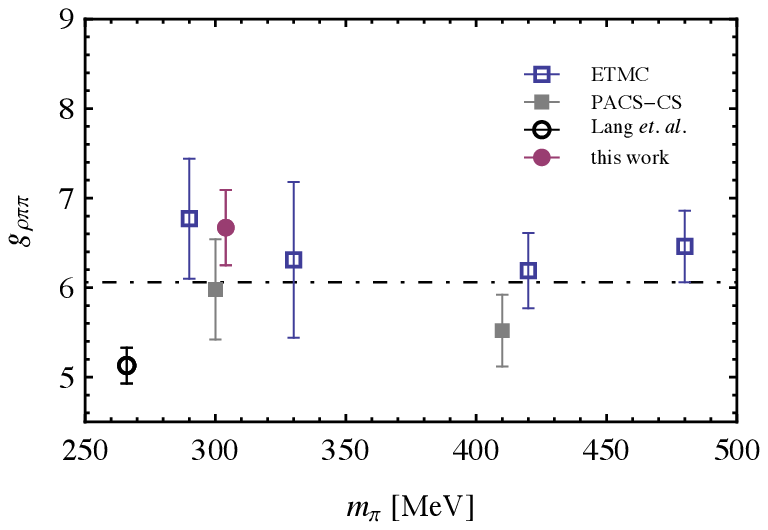}
\caption{Left panel: $\mr$ versus $m_\pi$ from recent lattice studies. We use
lattice units to remove the uncertainty associated with setting the scale. The band represents the
prediction from unitarized~$\chi$PT~\cite{Pelaez:2010fj} (note that this is not
a fit to our data).
Right panel: coupling as a function of $m_\pi$ from recent lattice studies. The
dot-dashed line is the PDG result~\cite{Nakamura:2010zzi}.
}
\label{fig:compare}
\end{figure*}

In Fig.~\ref{fig:compare} we  compare our results to other
recent lattice studies \cite{Feng:2010es,Lang:2011mn,Aoki:2006xj}. 
Our result for the coupling constant is in good agreement with other studies and
was determined with similar precision.
The value of $\mr$ is also compatible with the value reported by the other studies,
but our error bar is significantly smaller.
We note that the results reported by Lang and collaborators~\cite{Lang:2011mn} 
differ significantly from our results and the ones of other lattice studies.
This is presumed to be due to the fact that the volume used in their study is too small. 

The phase shifts and the values we obtained for $\mr$ and $\gpp$ are also in 
good agreement with one-loop unitarized chiral perturbation theory~\cite{Pelaez:2010fj}. 
To show the agreement, in the left panel of Fig.~\ref{fig:phaseplots} we plot the 
phase shift expected from unitarized~$\chi$PT to $O(p^4)$ for $m_\pi=304$~MeV. Note that this 
is not a fit to our data -- the only input from our study to this expression is the 
pion mass. Another check is presented in the left panel of Fig.~\ref{fig:compare} where
we plot the expectation for the mass of the $\rho$ as a function of the pion mass.
Our datapoint falls directly on top of the predicted band in the region where the 
unitarized~$\chi$PT prediction has the smallest uncertainty. This is not a trivial 
agreement since this band has no input from our study.

\section{Conclusions}\label{sec:conclusions}

We presented the first study of the $\rho$ meson decay using asymmetrical lattices. 
The calculation was carried out using nHYP-smeared clover fermions with two mass-degenerate 
quark flavors at a pion mass of $\mpi=304(2)$~MeV. We
used three ensembles of 300 dynamical gauge configurations and computed
the P-wave scattering phase shifts for the isospin $I=1$ channel.
Fitting the phase shifts using the effective range formula,
we found a coupling constant $\gpp=6.67(42)$ and resonance mass $\mr=827(3)(5)$~MeV. 
Using the computed value for the coupling constant and
assuming that it does not change significantly as the quark mass is lowered towards the
physical point, we estimate the physical decay width to be $\Gamma_{ph}=184(23)$~MeV, 
which is compatible with the experimental value $\Gamma_{exp}=149.1(8)$~MeV.

To compute the scattering phase shifts, we chose lattices volumes with one 
spatial direction elongated. By varying the elongation parameter, we were able 
to compute a set of scattering phase shifts evenly distributed throughout the 
spectral region where the $\rho$ decays. While the moving frame
formalism has been successfully used to study the $\rho$ decay for as low as $\mpi=266$ MeV
on cubic lattices, the finer momentum control offered by asymmetrical lattices will be needed to
study narrower resonances. 

We found that both the $\pi\pi(t)$ and $\rho(t)$ interpolators couple
to the two-pion scattering states and can be used to extract the ground state. 
However, using the variational method allows both the ground and first 
excited state to be computed and gives a more precise determination of the ground state. 

To compute the quark diagrams required for the two-pion operator, we had to employ 
stochastic methods. When employing spin-color dilution for the noise, we found that 
the gauge variance of the estimators is substantially larger than the noise variance.
It is then sufficient to use only two independent noise vectors per configuration --
additional vectors will add significantly to the analysis cost without reducing the
final error bars.

To compute the resonances parameters, we used the effective range formula. 
While this formula described the overall phase shift profile
well, the confidence level for our fit was only $Q\simeq8 \%$. We find that the
effective range formula cannot reliably describe the phase shifts for
the entire momentum range used in our study. However, the data points in the
resonance region are well described by this formula.

Our results are compatible with most recent lattice studies and our error bars
are comparable or better. Moreover, our results were in very good agreement 
with the predictions of unitarized chiral perturbation theory.

For future studies, we plan to increase the size of the variational basis as 
proposed by \cite{Prelovsek:2011im}, include smeared quarks, and compute the 
resonance parameters at a pion mass 
$\mpi\simeq200$~MeV. The systematic effects due to lattice spacing, finite volume, and
the inclusion of strange quark dynamics will be determined in further studies.

\begin{acknowledgments}

We would like to thank Frank Lee for suggesting this project and constant
encouragement, Anna Hasenfratz for providing us the codes for 
generating nHYP dynamical configurations, Sinya Aoki for 
useful discussions, and Guillermo Rios and Jose Pelaez for providing us with 
the unitarized $\chi$PT data.
This work has been performed on the IAC cluster
supported by the National Center for Supercomputing Applications, 
the HPCL clusters at The George Washington University, 
and the Keeneland Computing Facility at the Georgia Institute of Technology, 
which is supported by the National Science Foundation under Contract OCI-0910735. 
This work is supported in part by the U.S. Department of Energy grant DE-FG02-95ER-40907
and NSF CAREER grant PHY-1151648.
\end{acknowledgments}

\begin{appendix}
\section{Numerical evaluation of $\mathcal{Z}_{lm}(s,q^2;\eta_2,\eta_3)$}
\label{sec:zeta-functions}

\begin{figure*}[!t]
\includegraphics[width=3.1in]{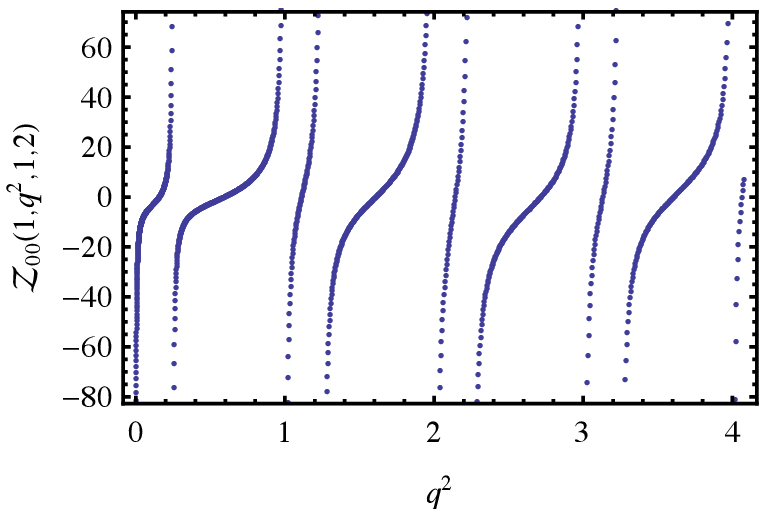}
\includegraphics[width=3.1in]{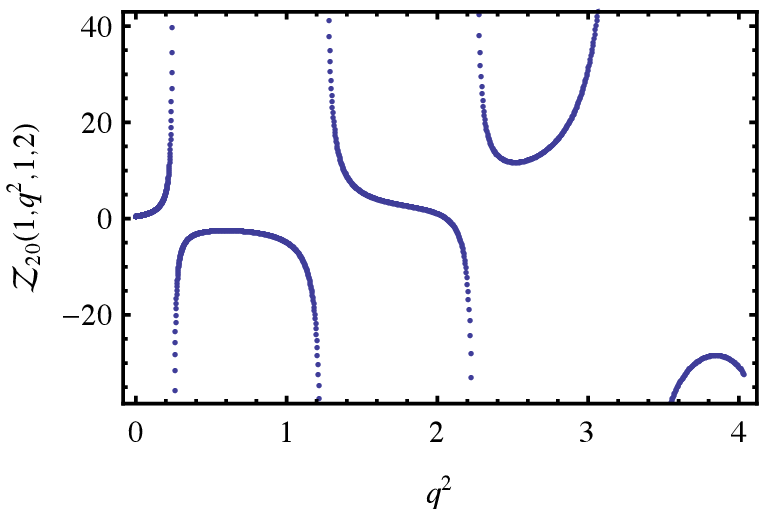}
\caption{$\calZ_{00}(1,q^2,1,2)$ and $\calZ_{20}(1,q^2,1,2)$.}
\label{fig:zetafig}
\end{figure*}

In the following we describe the numerical evaluation of the generalized zeta functions defined by
\be{
 \label{eqn:appendix_zeta_def}
 \mathcal{Z}_{lm}(s,q^2;\eta_2,\eta_3)=
 \sum_{\bn\inZ} {\mathcal{Y}_{lm}(\tilde{\textbf{n}})
 \over (\tilde{\textbf{n}}^2-q^2)^s}\;
}
for
\be{
\mathrm{Re}\, 2s < l+3\;.
}
We use the same strategy originally proposed by L\"usher in \cite{Luscher:1990ux}. Its
extension to asymmetrical lattices is given in \cite{Li:2003jn}. In the following,
we give the relevant formulas, describe the numerically procedure used in
this work, and give a list of values for comparison.

We begin by introducing the heat kernel
 \ba{
 \calK(t,\bx)&=&{1\over (2\pi)^3}\sum_{\bn\inZ} e^{i\tn\cdot\bx-t\tn^2} \label{eqn:hk1}\\
 &=& {\eta_2\eta_3\over (4\pi t)^{3/2}}
 \sum_{\bn\inZ} e^{-{1\over 4t}\left(\bx-2\pi\nhat\right)^2}\label{eqn:hk2}\;
 }
where $\nhat = (n_1,\eta_2\,n_2, \eta_3\, n_3)$ and  $\ntilde = (n_1,n_2/\eta_2, n_3/\eta_3)$. The first form is the definition of the heat kernel. Eq.~(\ref{eqn:hk2}) follows from Eq.~(\ref{eqn:hk1}) by applying the Poisson summation formula
\be{
\sum_{\bn\inZ}f(\bn)=\sum_{\textbf{k}\inZ}\int_{-\infty}^{\infty}f(\textbf{x})e^{i2\pi \dotp{k}{x}}d^3x\;
}
and will be useful for small times $t$. The derivatives of the heat kernel are defined by
\be{
 \calK_{lm}(t,\bx)=\calY_{lm}(-i\nabla_{\bx})\calK(t,\bx)\,
}
where the expression $\calY_{lm}(-i\nabla_{\bx})$ is to be understood as the replacement of the homogeneous polynomial
\be{
\calY_{lm}(\textbf{x})=x^lY_{lm}(\Omega_{\bx})\,,
}
using the rule $x_i\rightarrow\frac{\partial}{\partial x_i}$.
Lastly, we define the truncated heat kernel and its derivatives as
\be{
\label{eq:K_Lambda}
\calK^\Lambda(t,\bx)=\calK(t,\bx)-
{1\over (2\pi)^3}\sum_{|\tn|\leq\Lambda}
e^{i\tn\cdot\bx-t\tn^2}\;,
}
and
\be{
\calK^\Lambda_{lm}(t,\bx)= \calY_{lm}(-i\nabla_{\bx})\calK^\Lambda(t,\bx)\,.
}
A valid representation of the generalized zeta function, analytically continued to the half plane $\mathrm{Re}\, 2s>1/2$, can be written in terms of the truncated heat kernel as
\ba{
\nonumber
&&\hskip-1cm\calZ_{lm}(s,q^2;\eta_2,\eta_3)\\
&=&\sum_{|\tn|\leq\Lambda}
{\calY_{lm}(\tn) \over (\tn^2-q^2)^s}
+{(2\pi)^3\over\Gamma(s)}\left\{
{\delta_{l0}\delta_{m0}\eta_2\eta_3\over (4\pi)^2 (s-3/2)}\right.
\nonumber \\
&+&\int^1_0 dt\, t^{s-1}\left[
e^{tq^2}\calK^\Lambda_{lm}(t,\bzero)
-{\delta_{l0}\delta_{m0}\eta_2\eta_3\over (4\pi)^2 t^{3/2}}
\right] \nonumber \\
&+&\left.\int^\infty_1 dt\, t^{s-1}
e^{tq^2}\calK^\Lambda_{lm}(t,\bzero)
\right\} \;
} 
where the cutoff $\Lambda$ is chosen such that $\Lambda^2>\mathrm{Re}\,q^2$.
In order to compute the scattering phase shifts, we need to evaluate
$\calZ_{lm}(1,q^2;\eta_2,\eta_3)$ numerically. In the following we
describe the procedure used in this work.

\begin{table}[!b]
\begin{ruledtabular}
\centering
\begin{tabular}{lcc}
$q^2$&$\calZ_{00}(1,q^2,1,2)$&$\calZ_{20}(1,q^2,1,2)$\\
\hline
 0.2 &    6.35553 & 5.62348 \\
 0.41 & -5.08222 & -2.90547 \\
 0.62 &  1.16621 & -2.22911 \\
 0.83 &  10.9025 & -2.85516 \\
 1.04 & -33.5287 & -5.64013\\
\end{tabular}
\caption{A few values for $\calZ_{00}(1,q^2,1,2)$ and $\calZ_{20}(1,q^2,1,2)$.}
\label{tab:zetavals}
\end{ruledtabular}
\end{table}

For the case of interest, $s=1$, we have 
\ba{
\nonumber
 &&\hskip-1cm\calZ_{lm}(1,q^2;\eta_2,\eta_3)\\
 &=&\sum_{|\tn|\leq\Lambda}
 {\calY_{lm}(\tn) \over (\tn^2-q^2)}
 +{(2\pi)^3}\left\{
- {\delta_{l0}\delta_{m0}\eta_2\eta_3\over 8\pi^2}\right.
 \nonumber \\
 &+&\int^1_0 dt\left[
 e^{tq^2}\calK^\Lambda_{lm}(t,\bzero)
 -{\delta_{l0}\delta_{m0}\eta_2\eta_3\over (4\pi)^2 t^{3/2}}
 \right] \nonumber \\
 &+&\left.\int^\infty_1 dt\,
 e^{tq^2}\calK^\Lambda_{lm}(t,\bzero)
 \right\} \;.\label{eqn:zeta1}
 }
In order to evaluate $\calZ_{lm}(1,q^2;\eta_2,\eta_3)$, the infinite sums 
appearing in $\calK^{\Lambda}_{lm}(t,\bf 0)$
must be truncated. In the integration region $[1,\infty]$, we use 
Eq.~(\ref{eqn:hk1}) to define $\calK_{lm}(t,\bf 0)$. In this case subsequent 
terms in the sum are suppressed by $\mathrm{Re}\, e^{-t(\tn^2-q^2)}$,
and the sum converges the slowest for $t=1$. In the integration region 
$[0,1]$, we use the alternative definition
of the heat kernel to define $\calK^{\Lambda}_{lm}(t,\bf 0)$. 
In this case one has to approximate $e^{tq^2}\calK_{lm}(t,\bf 0)$. 
Again, the sum converges the slowest at $t=1$ with successive terms being 
suppressed by $\mathrm{Re}\, e^{-\frac{(\pi \hat{\bf n})^2}{t}+tq^2}$. The 
cutoff is then chosen such that the integrands are approximated to a desired 
precisions for $t=1$. In Table \ref{tab:zetavals} we list some values
for comparison, and in Figure \ref{fig:zetafig} we plot $\calZ_{00}(1,q^2;1,2)$ 
and $\calZ_{20}(1,q^2;1,2)$ in the range $0<q^2<4$.

\section{Determination of the lattice spacing}\label{app:lattice_spacing}

To determine the lattice spacing, we use the Sommer scale \cite{Sommer:1993ce}.
We work in the Coulomb gauge and extract the heavy quark potential from the relation
\be{
\avg{W_T^\dag(\bx,t)W_T(\bx+\textbf{r},t)}_U\approx c e^{-V(\textbf{r})T}\;,
}  
which is valid for large $T$ with Wilson line $W_T$ defined by
\be{
W_T(\bx,t)=\prod_{i=0}^{T-1}U_4(\bx,t+i)\;.
}
To fit the data, we use the functional form
\be{
V(r)=C+\frac{B}{r}+\sigma r+\lambda\p{\left.\frac{1}{r}\right|_{latt}-\left.\frac{1}{r}\right|}\label{eqn:staticpot}\;
}
where C is part of the quarks' self-energy, $\sigma$ is the string tension, $B=-\frac{3}{4}\alpha_s$,
and the last term is the difference between the lattice and continuum potentials. The Sommer
parameter is then determined through
\be{
r_0/a= \sqrt{\frac{1.65+B}{\sigma}}\;.
}
Using a combined fit to all three ensembles and averaging over all space-time points, we find
\be{
C=0.919(4)\;,\; B=-0.430(5)\;,\;\sigma=0.077(1),\;
}
and
\be{
r_0/a=3.984(21)\;.
}
To fix the lattice spacing, we take $r_0=0.5$ fm and obtain $a=0.1255(7)$ fm.

\end{appendix}

\bibliographystyle{jhep-3}
\bibliography{citations}
\end{document}